\documentclass[onecolumn,noshowpacs,pre,preprintnumbers,superscriptaddress,amsmath,amssymb,floatfix]{revtex4}
\usepackage[caption=false]{subfig}
\usepackage{graphicx}
\usepackage{color}
\usepackage{placeins}
\usepackage[normalem]{ulem}
\usepackage{braket}

\begin{document}
	
    \title{Percolation on spatial anisotropic networks}
	\author{Ouriel Gotesdyner}
	\affiliation{Department of Physics, Bar-Ilan University, Ramat Gan, Israel}
	\author{Bnaya Gross}
	\affiliation{Department of Physics, Bar-Ilan University, Ramat Gan, Israel}
	\author{Dana Vaknin Ben Porath}
	\affiliation{Department of Physics, Bar-Ilan University, Ramat Gan, Israel}
	\author{Shlomo Havlin}
	\affiliation{Department of Physics, Bar-Ilan University, Ramat Gan, Israel}
	\date{\today}
		
	\begin{abstract}
		\begin{center}
	    In Honour of Robert M. Ziff's 70th Birthday
	    \end{center}
        Many realistic systems such as infrastructures are characterized by spatial structure and anisotropic alignment. Here we propose and study a model for dealing with such characteristics by introducing a parameter that controls the strength of the anisotropy in the spatial network. This parameter is added to an existing isotropic model used to describe networks under spatial constraints, thus generalizing the spatial model to take into account both spatial and anisotropic features. We study the resilience of such networks by using a percolation process and find that anisotropy has a negative impact on a network's robustness. In addition, our results suggest that the anisotropy in this model does not affect the critical exponent of the correlation length, $\nu$, which remains the same as the known $\nu$ in 2D isotropic lattices.
	\end{abstract}

	\maketitle
		
	\section{Introduction}
    
    Recent years have seen a growing interest in network science in order to understand the structure and function governing many real complex systems. 
    Examples of such systems appear in every context of our everyday lives, epidemic spreading \cite{epi_pastor2015epidemic,epi_cohen2003efficient}, biological networks \cite{bioExample_barabasi2004network, biologicalFromDana_milo2002network}, climate networks \cite{climateFromDana_donges2009backbone, ClimateFromDana2_yamasaki2008climate}, traffic \cite{traffic,trafficExampleFromShlomo2_hamedmoghadam2021percolation}, economy \cite{EconomicsFromDana_bonanno2004networks,EconomiExampleFromShlomo2_smolyak2018interdependent}, the Internet \cite{internetFromDana_faloutsos2011power,InternetExampleFromShlomo2_carmi2007model}, social networks \cite{SocialFromDana_watts1998collective, SocialFromDana2_girvan2002community} and many more. Thus, understanding the behavior of such networks is an important step towards better understanding the reality of our lives \cite{intro_albert2002statistical, intr_boccaletti2006complex, intro_newman2006structure}.
    
    We regard networks in this study in an abstract manner --- structures composed of nodes and edges between them. For example, in a social network, the nodes represent people and the edges (links) represent their relations.  In a computer network like the Internet, nodes represent computers and links represent the cables connecting them. In order to understand the network's resilience we perform a percolation process, a process which has been well studies since it's introduction in 1941 \cite{ancientExampleFromShlomo}. In this process we start from a fully active network, where all nodes are connected, and randomly remove $1 - p$ fraction of the nodes ($p$ ranging from 0 to 1). A percolation phase transition occurs with the breaking down of the giant connected cluster, $P_\infty$, which is the largest connected component and of the order of the system's size. We denote the critical threshold  \textbf{}$p_c$ as the $p$ in which the giant component collapses. This process is used to study a network's resilience to random failures or attacks and provides important insights concerning it's robustness \cite{cohen2011resilience, network_resi_callaway2000network, RobustnessExampleFromShlomo1_cohen2010complex, RobustnessExampleFromShlomo2_Newman2010}. 

	Various network structures have been proposed and studied as models for different systems --- From the Erdos-Renyi's model which was introduced back in 1959 \cite{erdos59a, erdos1960evolution, gilbert1959random, er_bollobas1985random}, through scale-free networks \cite{scalefree_barabasi1999emergence}, multilayer networks \cite{multilayer_boccaletti2014structure}, multiplex networks \cite{multiplex_radicchi2017redundant}, interdependent networks \cite{interdependant_buldyrev2010catastrophic, interdependent_danziger2014introduction, interdependant_baxter2012avalanche, NoN_gao2011robustness} and networks of networks \cite{NoN_gao2011robustness, Non_leicht2009percolation} --- All of them proposing a variety and a growing complexity of the connections between the particular nodes, in one or more networks. Recent studies have introduced spatial constraints to the distribution of edge lengths \cite{Micha_zetaModel,danziger2020faster, BnayaAdd1_bonamassa2019critical, BnayaAdd2_gross2021interdependent, BnayaAdd3_gross2017multi, BnayaAdd4_vaknin2017spreading, Zeta_Model1_waxman1988routing, Zeta_Model2_bradde2010critical, ShlomosPowerlaw1_daqing2011dimension, ShlomosPowerlaw2_li2011percolation} helping us better understand realistic networks in which the link lengths are spatially limited. The spatial constraints are enforced in the model by using an exponential distribution for the edge lengths as suggested by Danziger et al \cite{Micha_zetaModel} based on realistic networks and similar to earlier studies made by Waxman \cite{Zeta_Model1_waxman1988routing} and Bradde et al \cite{Zeta_Model2_bradde2010critical}. By choosing an exponential distribution for the edge lengths (see Eq. (\ref{zeta equation})),  the strength of spatiality is controlled by a parameter $\zeta$, representing the typical link length in the network. This model helps to observe the effects of varying spatiality on the network's robustness \cite{Micha_zetaModel}. Note that the effects of a power law distribution of edge lengths in lattices,  has been studied by Daqing Li et al \cite{ShlomosPowerlaw1_daqing2011dimension, ShlomosPowerlaw2_li2011percolation}.
	
	However, spatiality alone is not enough in order to model many realistic network. As discussed by Vaknin et al \cite{ani_justi_vaknin2021cascading} many realistic networks are anisotropically aligned. For example, urban infrastructures such as power grids and traffic networks in coastal cities will bear strong anisotropic inclination parallel to their respective coastline. Further examples of anisotropic alignment can be witnessed in many kinds of networks - biological networks \cite{anisotropic_example_protein_eyal2006anisotropic}, superconductors \cite{anisotropic_example_super_conductors_gurevich1994anisotropic}, liquid crystals \cite{Liquid_Crystals_kelly1995anisotropic}, climate \cite{Ani_exapmle_climate_gursoy2017bioinspired} and more.
	
    Here, we study the robustness of networks that are both spatial and anisotropic using a model recently introduced in \cite{perez2021cascading} in the context of overload analysis. This allows us to study the impact of structural anisotropic characteristics on random failures of nodes in a spatial network. We introduce a parameter $\sigma$ that controls the anisotropy of the network and find that strong anisotropy causes the system, although having the same average degree, to become significantly less robust, pushing the percolation threshold higher for stronger anisotropy of the system (lower $\sigma)$. Our results also suggest that anisotropy, while affecting the length of the correlation, $\xi$ of the networks, it does not affect their scaling as expressed by an unchanging value of the correlation critical exponent $\nu$.
	
	\begin{figure}[ht!]
        \centering
            \subfloat[]{\includegraphics[width=0.5\linewidth]{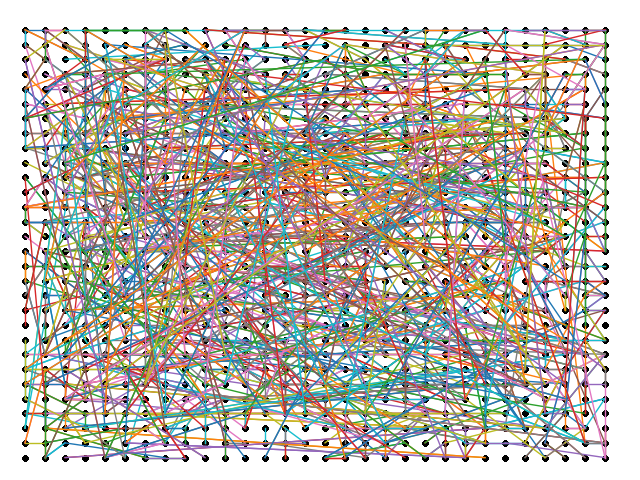}}
            \subfloat[]{\includegraphics[width=0.5\linewidth]{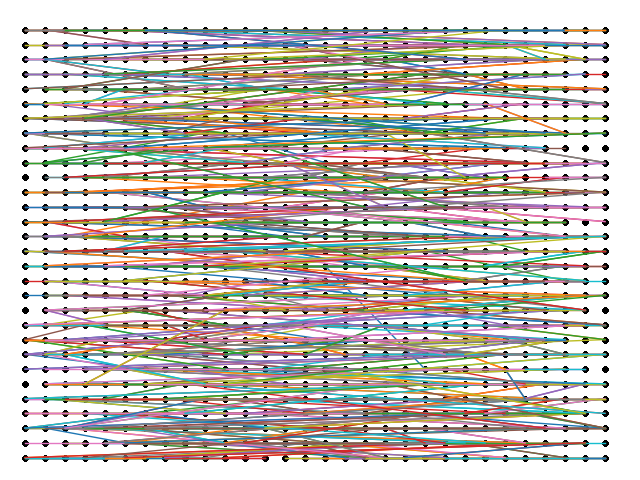}}\\
        	\subfloat[]{\includegraphics[width=0.5\linewidth]{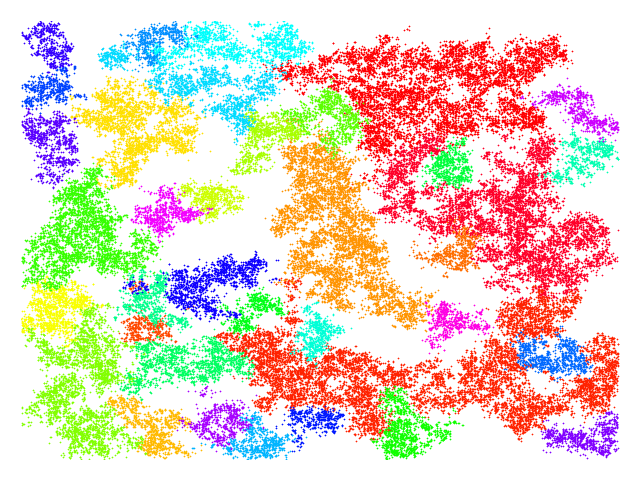}}
        	\subfloat[]{\includegraphics[width=0.5\linewidth]{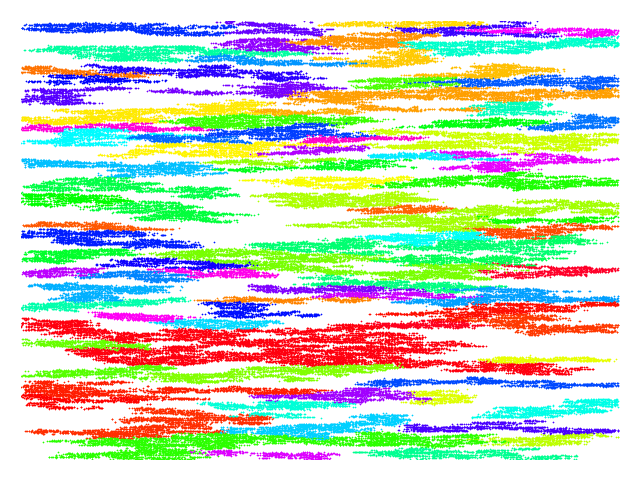}}\\
        \caption{
            \textbf{Illustration of the spatial anisotropic model. } (a), (b) Demonstration the edges layout in a small network of $L = 30$. We use the parameter $\sigma$ to control the anisotropic strength, see Eq. (\ref{sigma equation}). The network in (a) is isotropic with $\sigma = 5$ and in (b) is anisotropic with $\sigma = 0.1$. (c), (d) Demonstration of the largest clusters in a network for $\sigma = 5$ and $\sigma = 0.1$, respectively, for $p$ slightly below the percolation critical threshold $p_c$. Each of the large clusters (mass $>$ 500) shown is represented by a different color. The shape of the clusters in the anisotropic case is prolonged along the horizontal axis, which is the preferred direction of the edges. For all figures $\zeta$ = 5, $\braket{k}$ = 4 and $L = 1000$ for (c) and (d).}
        \label{fig:exampleClusters}	
    \end{figure}

	\section{Method}
	
    In our model we generate a network of $N$ nodes where the nodes are the sites of a two-dimensional square lattice. The parameter $L$ is the Euclidean linear size of the network, $N = L^D$ (where D is the dimension, in our case D = 2). We incorporate the spatial constraints by using a parameter $\zeta$ to control the spatial distribution of the length of the edges (see Eq. (\ref{zeta equation})), as proposed by Danziger et al \cite{Micha_zetaModel}. We generalize this spatial model to include anisotropic structures by introducing a new parameter $\sigma$ which controls the anisotropy of the system (see Eq. (\ref{sigma equation})). The edges are added between pairs of nodes having a defined average degree of $\braket{k}$. This is done by assigning a total number of $E = \frac{\braket{k}N}{2}$ edges to randomly selected nodes, resulting with a Poissonian degree distribution. An edge's Euclidean length, $r = \sqrt{dx^2 + dy^2}$, is chosen from the probability $P(r)$ between two nodes separated by a distance $r$, where
    \begin{equation}
    P(r) = \frac{e^{-r/\zeta}}{\zeta}  .
    \label{zeta equation} 
    \end{equation}
    Here, $\zeta$ represents the characteristic edge length in the network. In our model we also control the angle $\theta$ which represents the deviation of the edge from the $x$ axis positive direction. The angle $\theta$ is taken from the normal distribution: 
    \begin{equation}
    P(\theta) = \frac{e^{\frac{-\theta^2}{2\sigma^2}}}{\sigma \sqrt{2\pi}}
    \label{sigma equation}
    \end{equation}
    where the parameter $\sigma$ controls the standard angular deviation from the horizontal axis. Thus $\sigma$ controls the strength of anisotropy. For large $\sigma$ the network is isotropic, see Figs. \ref{fig:exampleClusters}(a) and \ref{fig:exampleClusters}(c), while for small $\sigma$ the network is anisotropic, see Figs. \ref{fig:exampleClusters}(b) and \ref{fig:exampleClusters}(d). In Figs. \ref{fig:exampleClusters}(c) and \ref{fig:exampleClusters}(d) we present examples of finite percolation clusters in the isotropic and anisotropic cases respectively. One can clearly see the difference in the typical cluster structure. In the isotropic case ($\sigma$ = 5) the clusters spread in both directions similarly. In contrast, the clusters in the anisotropic case ($\sigma$ = 0.1) are stretched along the horizontal axis.
    
    To simulate the percolation process we applied the algorithm proposed by Newman and Ziff \cite{newman2000efficient,newman2001fast} and the modifications made by Danziger et al \cite{danziger2020faster} for the calculation of the correlation length using disjoint sets. 
    
	\section{Results}

    \begin{figure}
		\centering
    		\subfloat[]{\includegraphics[width=0.5\linewidth]{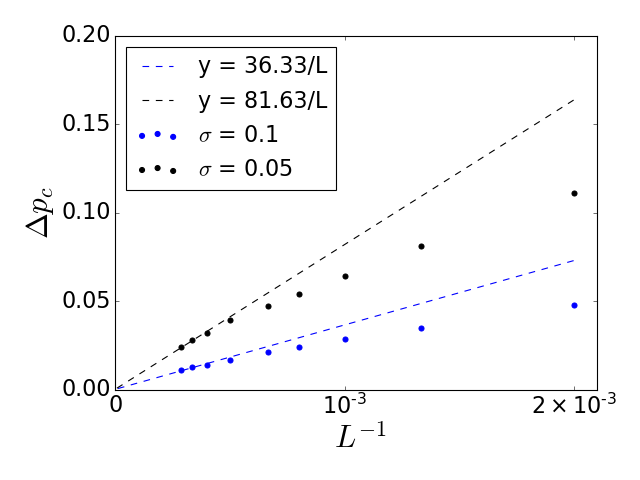}}
    		\subfloat[]{\includegraphics[width=0.5\linewidth]{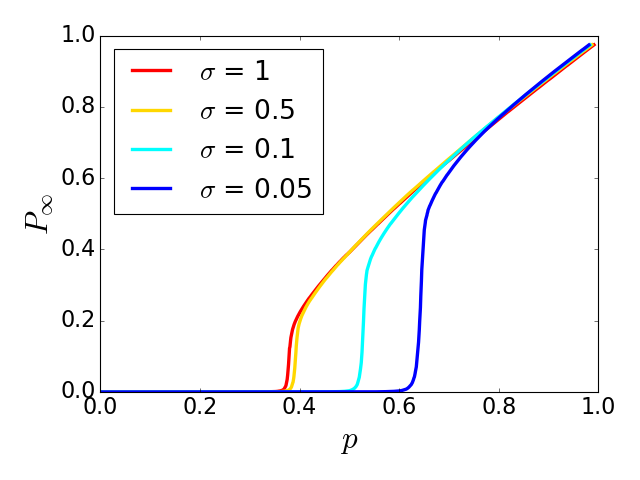}}\\   
            \subfloat[]{\includegraphics[width=0.5\linewidth]{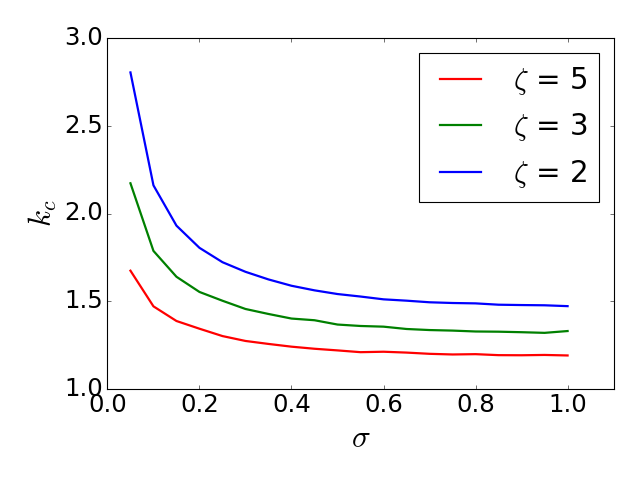}}
    		\subfloat[]{\includegraphics[width=0.5\linewidth]{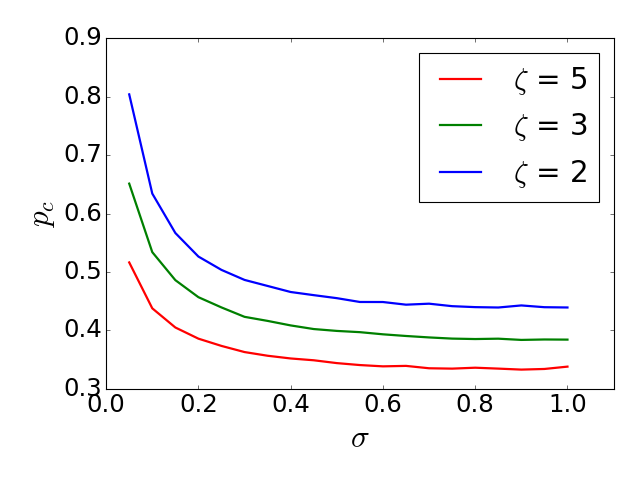}}
		\caption{
			\textbf{Percolation phase transition for different anisotropic strengths, $\sigma$. }(a) To reject the hypothesis that due to the anisotropy there exist different $p_c$ for spreading along the $x$-axis ($p_{cx}$) and along the y-axis ($p_{cy}$) we measured $\Delta p_c$ which is defined as the difference between $p_{cx}$ and $p_{cy}$. As the figure suggests the difference seems to diminish as the system size increases when extrapolating to $L$ = $\infty$, highly indicating that this difference is a finite size effect. Thus, we assume the same single $p_c$ for both axes even for the anisotropic case. The simulations for (a) were made with $\zeta = 2$ and $\braket{k} = 4$.
			(b) The relative size of the giant component, $P_\infty$, as a function of $p$, the larger is $p_c$ the more vulnerable the system is, since less nodes are needed to fail in order to break the system. This shows that isotropic networks (higher $\sigma$) are significantly more robust than anisotropic ones, although all are with the same average degree. The simulations for (b) were made with $\zeta = 3$ ,$L$ = 2000 and $\braket{k} = 4$.
			The critical thresholds (c) $k_c$ and (d) $p_c$ as a function of $\sigma$. One can see a similar behavior for both. Higher $p_c$ and $k_c$ are obtained for lower $\sigma$, implying that the anisotropy weaken the system dramatically to random failures. Note that $p_c$ and $k_c$ converge to a constant value as the system becomes isotropic enough. Simulations are shown for $\zeta = 3$ and $L$ = 2000 and for (d) $\braket{k} = 4$.}
		\label{fig:PhaseTransition}
	\end{figure}
	
	\begin{figure}
		\centering
    		\subfloat[]{\includegraphics[width=0.5\linewidth]{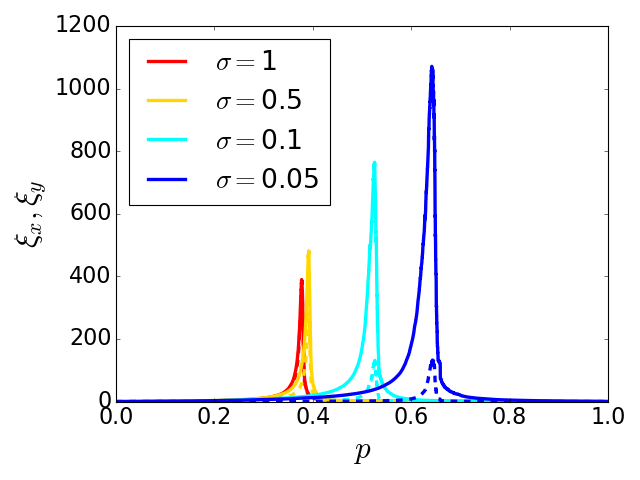}}
    		\subfloat[]{\includegraphics[width=0.5\linewidth]{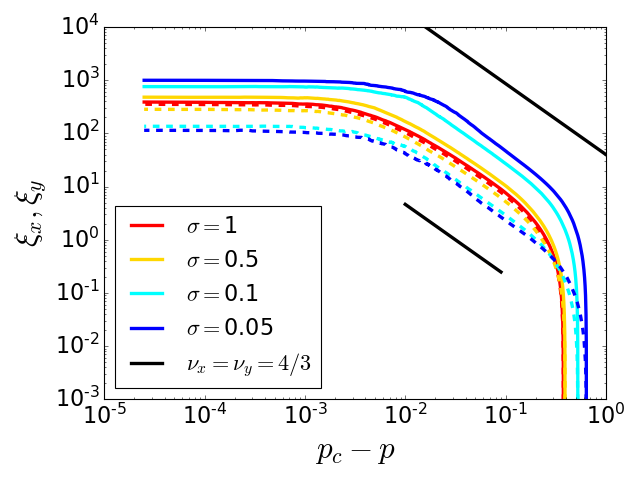}}\\
    		\subfloat[]{\includegraphics[width=0.5\linewidth]{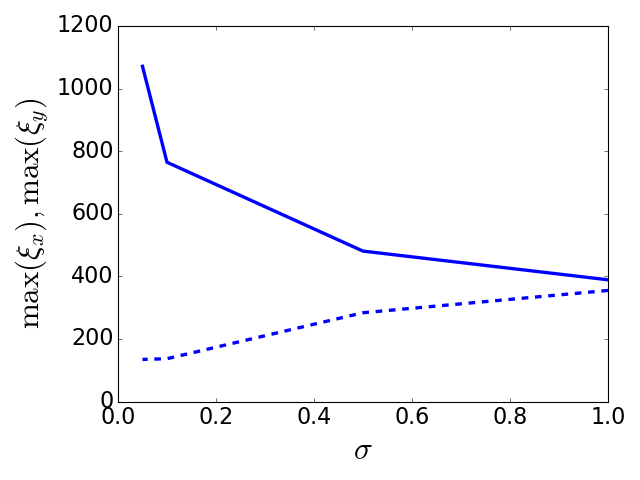}}
    		\subfloat[]{\includegraphics[width=0.5\linewidth]{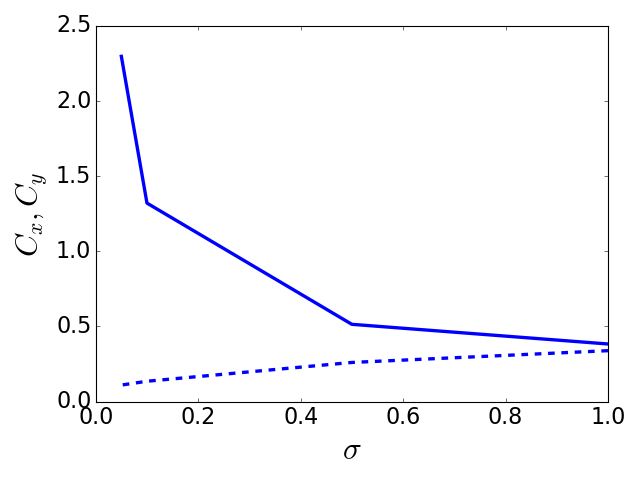}}
		\caption{
			\textbf{Correlation length for varying $\sigma$}. (a) Shows the correlation length, $\xi_x, \xi_y$ as a function of $p$ for different values of $\sigma$. The smooth lines represent $\xi_x$ and the dashed lines represent $\xi_y$. Maximal correlation length (the peaks) is obtained at the critical threshold $p_c$. (b) The correlation length exponent $\nu$ is given by $\xi = C|p-p_c|^{-\nu}$. To extract $\nu_x,\nu_y$ we measure the increase of the correlation length around $p_c$ in double logarithmic scales. The parallel curves indicate the similarity of the slopes, i.e., exponents, for the different values of $\sigma$ for both $\xi_x$ and $\xi_y$ suggesting that $\nu_x$ = $\nu_y$ as well as the unchanged $\nu$ under anistoropy. (c) Shows the maximal correlation length (the values of the peaks shown in (a)) and the convergence of maximal $\xi_x$ and $\xi_y$ to the same value for high $\sigma$. (d) Shows the significant difference in the coefficients $C_x$ and $C_y$ for the varying $\sigma$ and the convergence to the same value for large $\sigma$. Note, that while $\nu$ does not change with anisotropy, due to universality, the prefactors $C_x$ and $C_y$ are changed dramatically. For all figures $L$ = 2000, $\braket{k}$ = 4 and $\zeta$ = 3.}
		\label{fig:Xi and C}
	\end{figure}
	
    As seen in the difference between Figs. \ref{fig:exampleClusters}(c) and \ref{fig:exampleClusters}(d), strong anisotropy causes the clusters to stretch along the preferred axis (in our case, the horizontal one) which leads us to ask whether or not the percolation phase transition happens at a different threshold $p_c$ for each axis in the anisotropic case. To test this, we define criticality in each axis, $p_{cx}$ and $p_{cy}$ separately, as the fraction of active nodes, $p$, where the two vertical (to the x-axis) or horizontal edges respectively connect for the first time. We denote the difference $\Delta p_c \equiv p_{cx} - p_{cy}$\, and show in the Appendix in Fig. \ref{fig:deltaPCsigma} that $\Delta p_c$ increases as $\sigma$ decreases. However, as shown in Fig. \ref{fig:PhaseTransition}(a), $\Delta p_c$ also diminishes the larger the network's length $L$ is, suggesting that it is merely a finite size effect which approaches to 0 for infinite systems. Thus, our results suggest that high anisotropic infinite size networks have a $single$ percolation phase transition in both directions for large enough systems.
    
    Assuming a single $p_c$ we then measured the relative size of the giant component, $P_\infty$ as a function of $p$ for different values of $\sigma$ as shown in Fig \ref{fig:PhaseTransition}(b). Next, we measured $k_c$ (the value of $\braket{k}$ below which the network collapses) and $p_c$  for different values of $\sigma$ as shown in Figs. \ref{fig:PhaseTransition}(c) and \ref{fig:PhaseTransition}(d) respectively. We obtain a similar behavior for both $p_c$ and $k_c$ as they both increase for smaller values of $\sigma$. Thus, our results show that an anisotropic system is less robust compared to the isotropic case, as expressed by the higher values of $k_c$ and $p_c$ for smaller values of $\sigma$. This result can be explained by the fact that in the anisotropic case the vertical direction is less connected and the network breaks vertically more easily due to failure along the vertical axis.
    
    After analysing the impact of anisotropy on a system's stability we analyze it's impact on the correlation length, $\xi$, which is defined by \begin{equation}
        \xi^2 = \frac{\sum_\mu m_\mu I_\mu}{\sum_\mu m_\mu^2},
    \end{equation} \label{eq:nu_full}where $m_\mu$ is the size of cluster $\mu$ and $I_\mu$ is the moment of inertia of this cluster, summing over all clusters in the network. The moment of inertia of a cluster $\mu$ is defined by
    \begin{equation}
        I_\mu = \sum_{i}^{m_\mu}(r_i - \overline{r}_\mu)^2,
    \label{inertia} \end{equation}where $r_i,\overline{r}_\mu$ are the coordinates of the node $i$ in cluster $\mu$ and the center of mass of cluster $\mu$ respectively. Furthermore, we know that around the critical threshold ($p \rightarrow p_c$) the correlation length is given by \cite{nuExponent1_bunde1996percolation, nuExponent2_stauffer2018introduction}\begin{equation}
        \xi(p) = C|p-p_c|^{-\nu},
    \label{eq:nu_exponent}
    \end{equation}
    allowing us to extract the critical exponent $\nu$. In order to study the effects of anisotropy on the correlation length we separated the moment of inertia of each of the axes by defining $I_x$ and $I_y$ for each direction, horizontal (x-direction) and vertical (y-direction) in the following manner \begin{equation}
        {\xi^2_{x,y}} = \frac{\sum_\mu m_\mu {I_{x,y}}_\mu}{\sum_\mu m_\mu^2},
    \label{xiaxy}
    \end{equation}where the axial moment of inertia of each cluster is defined by \begin{equation}
        I_{q_j} = \sum_{i}^{m_\mu}({q_j}_i - \overline{q_j}_\mu)^2
    \label{inertiaxy}
    \end{equation}$q_j \in \{x,y\}$. As shown in Fig. \ref{fig:Xi and C}(a) and \ref{fig:Xi and C}(c) we find that the maximal correlation length becomes the same ($\xi_y \rightarrow \xi_x$) for the isotropic case. For the anisotropic case the maximal correlation lengths are very different, the correlation in y-direction is significantly smaller than that of the x-direction.  However, despite the difference between the correlation lengths of both directions in the anisotropic case, we notice for both x and y directions, a similar slope in Fig. \ref{fig:Xi and C}(b), that is a similar characterizing criticality --- suggesting the same value of the critical exponent $\nu_x = \nu_y$. Fig \ref{fig:Xi and C}(b) has been derived from the relation around the critical threshold in Eq. (\ref{eq:nu_exponent}), as shown in Fig. \ref{fig:Xi and C}(b). Thus, our results suggest that while the exponents are the same, the prefactors $C$ of Eq. (\ref{eq:nu_exponent}) are different in the anisotropic case, for x and y directions. Indeed, to complete the picture we calculated the different coefficients $C_x$ and $C_y$ for the different values $\sigma$ as shown in Fig. \ref{fig:Xi and C}(d). 
    
 	\section{Discussion and Conclusion}

 	It seems clear, based on our results, that isotropic networks are more robust than anisotropic ones. Lower values of $\sigma$ causing for higher $k_c$ and $p_c$ and as we increase $\sigma$ they decrease. Our results also suggest that above a certain value of $\sigma$, around $\sigma$ = 0.5, these critical thresholds seem to converge to a fixed value (which is different for each $\zeta$, as shown in Figs. \ref{fig:PhaseTransition}(c) and (d). Furthermore, we have seen that the scaling of the correlation length near criticality, $\nu_x, \nu_y$, for the $x$ and $y$ axis respectively, are equal and unaffected by the anisotropy --- implying that the prefactors such as $C_x$ and $C_y$ are strongly affected by the anisotropy, see Figs. \ref{fig:Xi and C}(c) and \ref{fig:Xi and C}(d).
 	
 	The fact that the correlation exponents are the same, $\nu_x = \nu_y$ and remain unchanged according to our simulations even for different strengths of anisotropy (different values of $\sigma$) differs from the results yielded by another anisotropic model proposed by Dayan et al \cite{Dayan}. This model was proposed for invasion percolation and involved different porosity values for the different layers in a 2d lattice. They introduce a parameter $\Delta$ which stands for that difference and had found via simulations that a larger $\Delta$ increase the difference between $\nu_x$ and $\nu_y$. The question why both models yield different behaviors is left open for further research.
 	
 	\section*{ACKNOWLEDGMENTS}

    This work was supported by the Israel Science
    Foundation, (Grant No. 189/19) and the joint China-Israel Science Foundation (Grant No.3132/19),  BSF-NSF, and DTRA (Grant No. HDTRA-1-19-1-0016) and the Bar-Ilan cyber security center. 
    
    \section*{Appendix A. Italian powergrid and anisotropic crossover}
    
    \renewcommand{\thefigure}{A\arabic{figure}}

    \setcounter{figure}{0}
    
\begin{figure}
    	\centering
        	\subfloat[]{\includegraphics[width=0.5\linewidth]{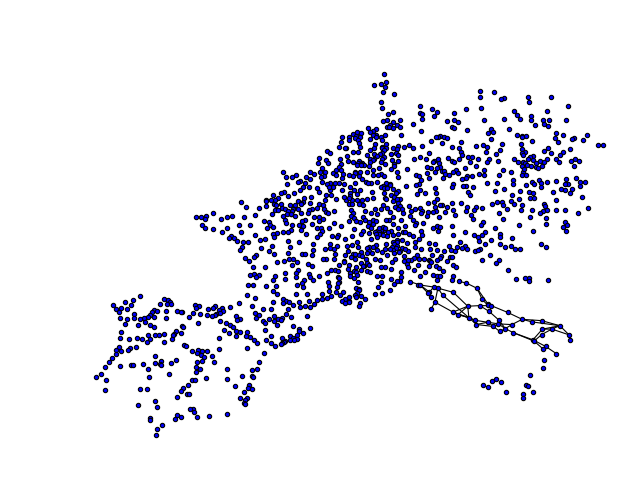}}
        	\subfloat[]{\includegraphics[width=0.5\linewidth]{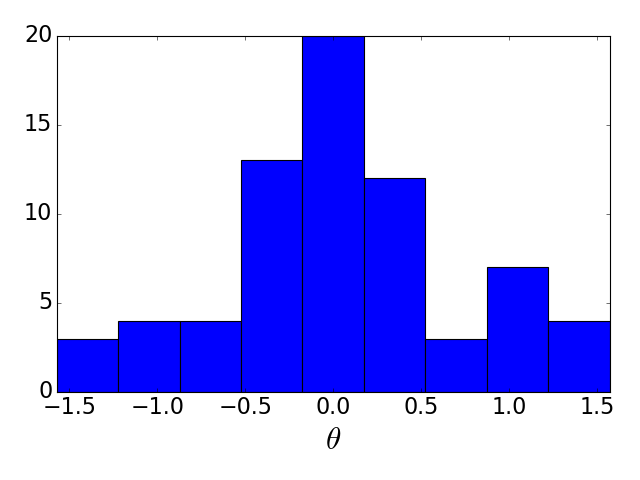}}\\
    	\caption{
     		\textbf{An example of a realistic anisotropy.} In (a) we see a depiction of the European power grid, focusing only on the Italian mainland (where edges are shown). (b) Is a histogram of the anglular deviation of the links from the horizontal axis after a rotation of 24 degrees. The standard deviation of the angular deviation yields that according to our model for mainland Italy $\sigma = 0.64$.}
     	\label{fig:Demonstration}
 	\end{figure}

    In this appendix we show an example of a real network, the powergrid of the Italian mainland, which demonstrates an anisotropic inclination, corresponding to a $\sigma$ value of 0.64 as shown in Fig. \ref{fig:Demonstration}.
    
		\begin{figure}
		\centering
    		\subfloat[]{\includegraphics[width=0.5\linewidth]{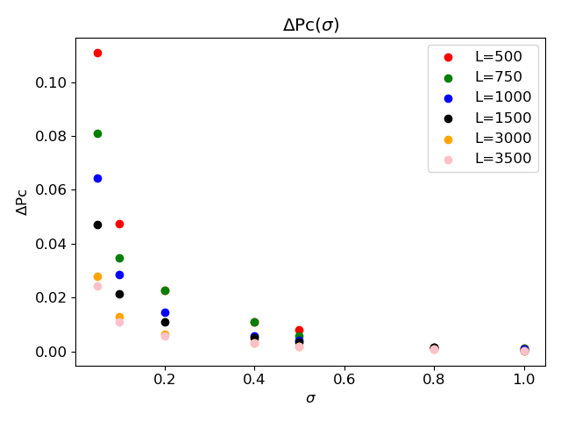}}
		\caption{
			\textbf{The difference $\Delta p_c$ = $p_{cx} - p_{cy}$ as a function of $\sigma$.} As shown in this figure, we obtain larger differences between the critical threshold for the $x$-axis $p_{cx}$ and the critical threshold for the $y$-axis $p_{cy}$ for stronger anisotropy of the network is (smaller $\sigma$). However, as discussed above according to the results obtained in Fig. \ref{fig:PhaseTransition}(a) we conclude that this difference is a finite size effect and approach to 0 in infinitely large anisotropic systems.}
		\label{fig:deltaPCsigma}
	\end{figure}
	
	\FloatBarrier
	\bibliographystyle{naturemag_4etal}
	\bibliography{sample}
	
\end{document}